\documentclass[prl,twocolumn,showpacs,amssymb]{revtex4}

\usepackage{graphics}
\usepackage{float,epsfig}
\begin{document}

\title{Long standing problem of $^{210}$Bi and the realistic 
neutron-proton effective interaction}

\author{L. Coraggio$^{1}$, A. Covello$^{1,2}$, A. Gargano$^{1}$,
and N. Itaco$^{1,2}$} 
\affiliation{$^{1}$Istituto Nazionale di Fisica Nucleare, 
Complesso Universitario di Monte S. Angelo, I-80126 Napoli,
Italy \\
$^{2}$Dipartimento di Scienze Fisiche, Universit\`a
di Napoli Federico II,
Complesso Universitario di Monte S. Angelo,  I-80126 Napoli,
Italy}
\

\date{\today}

\begin{abstract}

The odd-odd nucleus $^{210}$Bi is studied within the framework of the  shell 
model using  effective two-body matrix elements derived  
from the CD-Bonn nucleon-nucleon potential. The experimental energies of the 
proton-neutron multiplet
$\pi h_{9/2} \nu g_{9/2}$ are remarkably well reproduced by the  
theory, which accounts for the $1^{-}$ state being the ground state instead of
the  $0^{-}$ predicted by the Nordheim strong coupling rule.
It is shown that the core-polarization effects are 
crucial to produce this inversion. 
The similarity between neutron-proton multiplets in the $^{132}$Sn 
and  $^{208}$Pb regions is discussed
in connection with the effective interaction.

\end{abstract}    

\maketitle

A fundamental goal  of nuclear physics is to understand
the properties of nuclei starting from the forces between nucleons. Within the
framework of the shell model, which is the basic framework for 
the study of complex nuclei in terms of nucleons, this implies the derivation
of the two-body 
effective interaction $V_{\rm eff}$ from the free
nucleon-nucleon ($NN$) potential.  This  has long been a main problem of
microscopic nuclear structure theory.
Suffice to say that  since the early 1950s  through the
mid 1990s in the vast majority of shell-model calculations either empirical 
effective interactions containing adjustable  parameters have been used  or 
the two-body matrix elements themselves have been treated as free parameters.
 
Great progress toward "realistic" shell-model calculations
has been achieved over the last decade, proving 
the ability of effective interactions derived from modern $NN$ 
potentials to provide, with no adjustable parameters, an accurate description of 
nuclear properties. This makes it very interesting to put to the test realistic 
shell-model calculations on peculiar nuclear properties which could not be 
unambiguously explained by earlier calculations. One such case 
is provided by the nucleus $^{210}$Bi.

The nucleus $^{210}$Bi with one proton and one neutron outside doubly magic
$^{208}$Pb core has long been the subject of theoretical studies. In this 
nucleus there are ten states below 600 keV with angular momenta from zero to 
nine, which in terms of the shell model  constitute a neutron-proton multiplet 
arising from the 
$\pi h_{9/2} \nu g_{9/2}$ configuration. The experimental determination 
of the ground-state spin of $^{210}$Bi, $J^{\pi}=1^-$, dates back to some 
fifty years ago 
\cite{Erskine62} and since then it has been a challenge to shell-model 
theorists to explain 
this feature, which is at variance with the prediction, $J^{\pi}=0^-$, of the
Nordheim strong coupling rule \cite{Nordheim50}. 

Early calculations \cite{Kim63,Mello63,Hughes66} made use of empirical 
two-body interactions containing adjustable parameters. These studies led 
to the conclusion that tensor-force components were needed to explain the 
inversion  of the $0^-$ and $1^-$ states. Realistic shell-model calculations 
starting from 
the Hamada-Johnston $NN$ potential \cite{Hamada62} 
were performed for $^{210}$Bi by  
Herling and Kuo \cite{Herling72} in the early 1970s. The hard core of the 
Hamada-Johnston potential was handled by using the Brueckner reaction 
matrix $G$. Then,  the effective 
two-body interaction was constructed by 
adding to  the $G$ matrix certain classes of diagrams arising from 
core-excitation processes, accounting in this way  for configurations left out 
of the space  in which the shell-model calculation was  performed.
A good overall agreement 
between the calculated and experimental spectra was obtained in 
\cite{Herling72}, but
the above inversion  was not explained. 
Some twenty years later, calculations in the $^{208}$Pb region were performed 
by Warburton and Brown
\cite{Warburton91} with  some modifications to the Kuo-Herling interaction.
In particular, for the $\pi h_{9/2} \nu g_{9/2}$  multiplet, modifications of 
all the corresponding two-body matrix elements were introduced to reproduce 
the experimental energies. 

In  the work of \cite{Ma73} the results of a shell-model calculation for nuclei in the 
lead region including
$^{210}$Bi were  presented together with a brief but comprehensive review of  studies on 
this nucleus  through 1972. The authors of \cite{Ma73} made use of an empirical effective interaction
with no tensor components but containing multipole forces with adjustable strengths to 
account for model-space truncation effects. Their conclusion was that an unambiguous explanation
of the inversion of the $0^-$ and $1^-$ states still remained to be given. In fact, they showed 
that their model was not able to simultaneously describe the correct ordering of the two states 
and the energies of the higher-lying states  of the $\pi h_{9/2} \nu g_{9/2}$ multiplet.
They pointed out that, while  the studies \cite{Kim63,Hughes66} evidenced 
the key role of the tensor force to reproduce the whole multiplet,
realistic effective interactions as that
used in \cite{Herling72} had failed to reproduce the  $^{210}$Bi ground state
although the starting $NN$ potential  contained  explicitly tensor components.  

 This situation had not changed much after more than two decades, 
as is reflected in the work of Ref. \cite{Alexa97a}, where 
a description of $^{210}$Bi is given by using  a phenomenological
model, which, besides an empirical neutron-proton effective interaction, includes
explicitly macroscopic degrees of freedom through a $^{208}$Pb vibrating core.  
  
As mentioned above, in the last ten years or so shell-model calculations employing realistic 
effective 
interactions derived from modern $NN$ potentials have produced results in 
remarkably  good agreement with experiment for a number of nuclei in 
various mass regions.
In the most recent calculations, the difficulty of dealing with  potentials 
having a  strong repulsive core
has been overcome  by resorting   
to the so called $V_{\rm low-k}$ approach \cite{Bogner02}. Within this approach 
a low-momentum $NN$ potential, $V_{\rm low-k}$, is obtained by 
integrating out the high-momentum modes of the $NN$ potential  down to a 
cutoff momentum $\Lambda$, with the requirement that the deuteron binding energy 
as well as the half-on-shell $T$ matrix  of the original potential are 
preserved.  
This $T$-matrix equivalence procedure, as described in detail in Ref. \cite{Bogner02}, 
produces a smooth potential
which is suitable for being used directly in nuclear structure
calculations.
The shell-model effective interaction is then constructed 
within the framework of a $\hat Q$-box  folded-diagrams expansion \cite{Kuo90}, 
namely it is expressed as a $\hat Q$-box, which is vertex function composed 
of irreducible valence-linked diagrams at any order in  $V_{\rm low-k}$, plus the 
folded-diagram  series. Note that the $G$-matrix   
calculation of the effective interaction performed by Herling and Kuo 
\cite{Herling72} is equivalent to retain the first three terms of 
the   $\hat Q$-box without folded diagrams, i.e., the $G$ matrix and  
the core-polarization diagrams corresponding to  one particle-one hole ($1p1h$)
and two particle-two hole ($2p2h$) excitations.

In recent years, particular attention from the experimental and 
theoretical point of view has been focused 
on nuclei around 
doubly magic $^{132}$Sn. In this region, the counterpart of $^{210}$Bi is the 
nucleus  $^{134}$Sb with one proton and one neutron in the $Z=50-82$ and 
$N=82-126$ shells, respectively. In this case, the 
ground state has $J^{\pi}=0^-$ and is nearly degenerate with the first-excited 
$J^{\pi}=1^-$ 
state, the latter lying at 13 keV. These two states are members of the lowest 
neutron-proton multiplet arising  from the $\pi g_{7/2} \nu f_{7/2}$ 
configuration. 
Very recently, we have performed a shell-model study of this nucleus 
\cite{Coraggio06} 
making use of an effective interaction derived from the 
CD-Bonn $NN$ potential \cite{Machleidt01}. 
Our results turned out to be in very good agreement with the experimental data 
including the very small energy spacing between the $0^-$ and $1^-$ states. 
A main finding of this study was that core polarization effects, in particular 
those arising from $1p1h$ excitations, 
introduce modifications in the effective interaction which are essential to 
reduce  the spacing between the $1^-$ and $0^-$ states. 

The above achievement is at the origin of the present paper, where we present 
the results of a realistic shell-model calculation for $^{210}$Bi conducted 
along the 
same lines as that for $^{134}$Sb. Note that each of the two single-particle 
levels composing
the $\pi g_{7/2} \nu f_{7/2}$ multiplet in  $^{134}$Sb has a counterpart 
with the same radial quantum number and one more unit in the angular momenta 
$l$ and $j$
in the $\pi h_{9/2} \nu g_{9/2}$  multiplet  of $^{210}$Bi.
As we shall see, this multiplet is very well reproduced with the $0^-$ and 
$1^-$ states in correct order. 

Our study  of  $^{210}$Bi has been performed 
assuming $^{208}$Pb  as a closed core and taking  as model space for the 
valence proton and neutron the six levels  of the 82-126 shell and the  
seven levels  of the 126-184 shell, respectively.
 
The two-body effective interaction $V_{\rm eff}$ has been derived by means of 
the  $\hat Q$-box  folded-diagrams expansion \cite{Kuo90}
from the CD-Bonn $NN$ potential, the short-range repulsion of the latter being 
renormalized by use of the  $V_{\rm low-k}$  potential \cite{Bogner02}.
As in our previous studies for nuclei around $^{132}$Sn 
\cite{Coraggio05,Coraggio06,Covello2},  the cutoff momentum  $\Lambda$ is given
the value 2.2 fm$^{-1}$.
This $V_{\rm low-k}$, with addition of the Coulomb interaction for
protons, is then used to calculate the 
$\hat{Q}$-box . In our calculation, we include one- and two-body diagrams up
to second order in the interaction, as they are explicitly shown in 
\cite{Shurpin83}.
The computation  of these diagrams 
is performed within the harmonic-oscillator basis using  intermediate 
states composed of all possible hole states and 
particle states restricted to the five proton and neutron shells
above the Fermi surface. This guarantees the stability of the
results when increasing the number of intermediate particle states. 
The oscillator parameter is $6.88$ MeV, as obtained  from the expression 
$\hbar \omega=45 A^{-1/3} -25 A^{-2/3}$. Once the $\hat{Q}$-box 
is calculated, the series of  
folded diagrams is summed up to all orders using the Lee-Suzuki ù
iteration method \cite{Suzuki80}. 

The effective interaction obtained by this procedure contains
one- and two-body contributions, the former determining, once summed to the 
unperturbed Hamiltonian, the single-particle energies. 
We have used the subtraction procedure of Ref.
\cite{Shurpin83} so as to retain only the two-body term $V_{\rm eff}$, while 
the single-particle energies 
have been taken from experiment.
The single-proton and -neutron energies adopted in our calculation are reported
in Table I. 
They are relative to $^{208}$Pb as obtained from the spectra of $^{209}$Bi and 
$^{208}$Pb \cite{NNDC} using mass excesses from \cite{Audi03}.

\begin{table}
\caption{Proton and neutron single-particle energies (in MeV).} 

\begin{ruledtabular}
\begin{tabular}{c|c|c|c}
$\pi(n,l,j)$&$\epsilon$ & $\nu (n,l,j)$& $\epsilon$\\
\colrule
$0h_{9/2}$ & -3.80 & $1g_{9/2}$ & -3.94\\
$1f_{7/2}$ & -2.90 & $0i_{11/2}$ & -3.16 \\
$0i_{13/2}$ & -2.19 & $0j_{15/2}$ & -2.51\\
$1f_{5/2}$ & -0.97  & $2d_{5/2}$ & -2.37\\
$2p_{3/2}$ &  -0.68 & $3s_{1/2}$ & -1.90\\
$2p_{1/2}$ &  -0.16 & $1g_{7/2}$ & -1.44\\
&   &                $2d_{3/2}$ & -1.40  \\
\end{tabular}
\end{ruledtabular}
\end{table}

Let us now come to the  results of our calculations, which  have been performed
by using the OXBASH shell-model code \cite{Oxbash}.
As for the binding energy of
the ground state,  our  calculated value is $8.347 \pm0.003$ MeV, which 
compares very well with  the experimental one,  
$8.404 \pm 0.002$ MeV \cite{Audi03}.  
Note that the error on the calculated value arises from the experimental errors
on  the proton and neutron separation energies of $^{209}$Bi and $^{209}$Pb 
\cite{Audi03}.

\begin{figure}[H]
\begin{center}
\includegraphics[scale=0.75,angle=0]{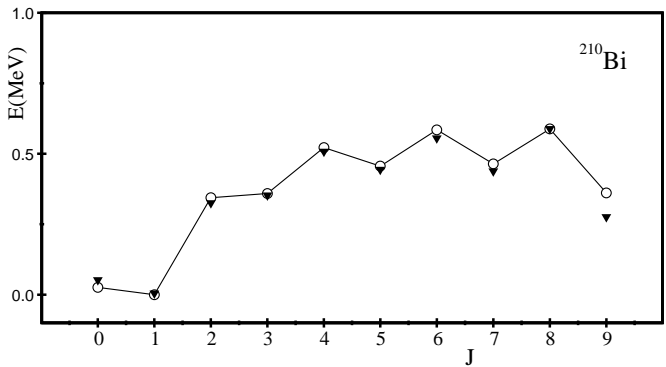}
\end{center}
\caption{Proton-neutron $\pi h_{9/2} \nu g_{9/2}$ 
multiplet in $^{210}$Bi. The theoretical results are represented by open 
circles while the experimental data by solid triangles. \label{fig1}}
\end{figure}

The calculated wave functions of the yrast states with $J^{\pi}$ from $0^-$
to $9^-$ are all dominated by the $\pi h_{9/2} \nu g_{9/2}$  configuration with
very little configuration mixing. In fact, the percentage of this component 
ranges from 96\% to 100\% except for the $1^-$ state where it has the minimum 
value of 91\%. These states represent therefore the ten members of  the 
$\pi h_{9/2} \nu g_{9/2}$
multiplet and their excitation energies are reported in Fig. \ref{fig1}, 
where they are compared with the   
experimental yrast states with the same spin and parity \cite{NNDC}.
Note that besides the ten members of the multiplet, only a  second $1^-$ 
state is found below 600 keV in both the experimental and calculated spectra. 

From Fig. \ref{fig1}, we   see that the experimental pattern is well reproduced by the 
theory. We correctly predict the energy decrease from the $0^-$ to the 
$1^-$ state as well as the sharp and slight increase occurring in the next two
steps. 
Then, from the $3^-$ state on the experimental and theoretical patterns 
stagger with the same magnitude and phase. The results shown in Fig. \ref{fig1} are also
in  remarkably  quantitative  agreement with the experimental data.
In fact, the discrepancies are less than  50 keV for all the states except  
the $9^-$, which 
is predicted to lie at 90 keV above the  experimental one.

Let us now discuss our  effective interaction. In our study of 
$^{134}$Sb \cite{Coraggio06} 
the effective two-body interaction was derived using precisely  the same approach  
as  that adopted here for   $^{210}$Bi and  it turned out that the $\pi g_{7/2} \nu f_{7/2}$ 
multiplet was very well described. In this context, we 
analyzed the various terms contributing to the effective interaction
in order  to understand their relative importance in determining 
the neutron-proton matrix elements  leading to a successful description of 
the multiplet.  This provided clear evidence of the crucial  role played 
by the $1p1h$ excitations especially in  regard to the $0^{-}-1^{-}$ spacing.

\begin{figure}[H]
\begin{center}
\includegraphics[scale=0.6,angle=0]{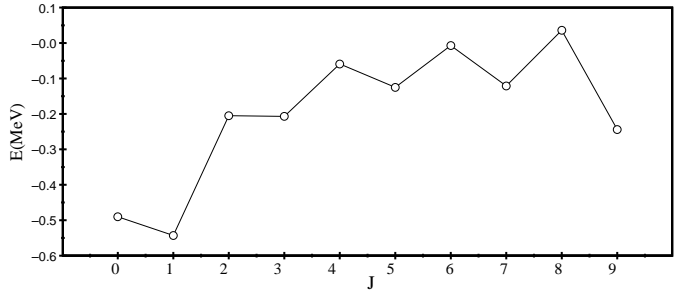}
\end{center}
\caption{Diagonal matrix elements of the two-body effective
  interaction for the $\pi h_{9/2} \nu g_{9/2}$
  configuration. \label{fig2}} 
\end{figure}

\begin{figure}[H]
\begin{center}
\includegraphics[scale=0.6,angle=0]{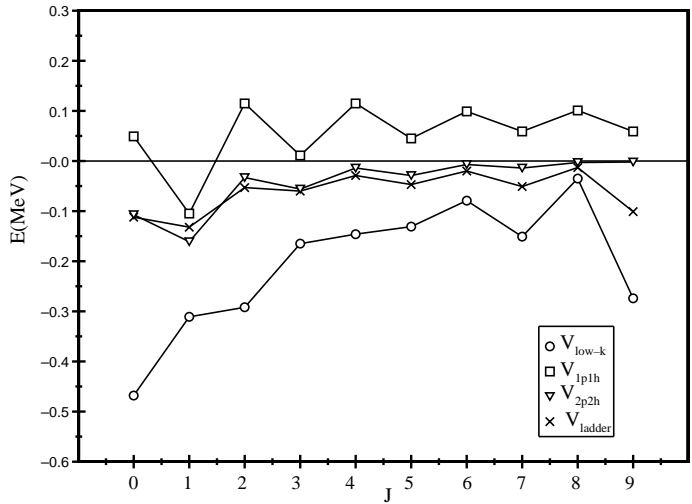}
\end{center}
\caption{Diagonal matrix elements of $V_{\rm low-k}$ and contributions
from the two-body second-order diagrams for the $\pi h_{9/2} \nu
g_{9/2}$ configuration. See text for comments. \label{fig3}}
\end{figure}

Here, we have performed the same analysis for the  $\pi h_{9/2} \nu g_{9/2}$  
multiplet in $^{210}$Bi.
In Fig. \ref{fig2}, we report the diagonal matrix elements of the effective interaction 
as a function of $J$ for the $\pi h_{9/2} \nu g_{9/2}$ configuration. 
As expected, we see that their behavior  is quite 
similar to that shown in Fig. \ref{fig1} for the energies of the  multiplet.
Then,  we show in Fig. \ref{fig3} the four  two-body second-order contributions to the 
effective interaction, namely
the corresponding matrix elements of $V_{\rm low-k}$ and  the
renormalizations due to core polarization through $1p1h$ and $2p2h$ 
excitations as well as to the ladder diagrams which account for excluded 
configurations above 
the chosen model space. As in the case of $^{134}$Sb, 
the folding contribution turns out to be quite irrelevant for the 
behavior of the neutron-proton  multiplet.
From Fig. \ref{fig3} it is clear that the features of the neutron-proton 
effective interaction evidenced in our study of $^{134}$Sb persist in the
$^{208}$Pb region.
In fact, we see that also in this region 
the neutron-proton  matrix elements of $V_{\rm low-k}$ do not follow the 
experimental
behavior, the most attractive one being that for $J^{\pi}=0^-$.
This behavior is significantly modified by the $V_{\rm 1p1h}$ term, which is
always repulsive except for $J^{\pi}=1^-$.

It is worth mentioning that in a preliminary calculation on $^{210}$Bi 
\cite{Covello04} the quality of agreement between theory and experiment was not
as good as that obtained here. In particular, we did not predict  the 
$0^-$ and $1^-$ levels in correct order.
However, also in that case the contribution from $1p1h$ excitations 
was essential to reduce the spacing between the $1^-$ and $0^-$ states, although
not sufficient to produce the inversion of 
these two states. It turned out that the 
correct ground state could be obtained increasing by a factor of about 2.5 
the diagonal matrix elements of $V_{\rm 1p1h}$ for the
$\pi h_{9/2} \nu g_{9/2}$ configuration.

In the present calculation, the $1p1h$  contribution brings about the right
effect without the introduction of any modification.
This is traced to the improvement in the calculation of the effective
interaction resulting from the increased number of intermediate states used 
in the computation of the $\hat Q$-box diagrams. 
The same result was also found  for  $^{134}$Sb \cite{Coraggio06}.

At this point, we may conclude that our
effective interaction contains the noncentral components needed
to give the $0^-$ and $1^-$ levels in correct order. These arise 
from  virtual interactions of the  core particles induced by the $NN$ potential and 
their weight with respect to the central 
components may  
be quite different from that relative to the  $NN$ potential.

In summary, we have shown that a shell-model interaction derived from a modern
$NN$ potential without any modification leads to a direct solution of the long
standing problem of $^{210}$Bi. 

\begin{acknowledgments}
This work was supported in part by the Italian Ministero
dell'Istruzione, dell'Universit\`a e della Ricerca  (MIUR).
\end{acknowledgments}


\begin{thebibliography}{9}

\bibitem{Erskine62} J. R. Erskine, W. W. Buechner, and H. A. Enge, Phys Rev. {\bf 128}, 720 (1962).
\bibitem{Nordheim50} L. W. Nordheim, Phys Rev. {\bf 78}, 294 (1950).
\bibitem{Kim63} Y. E. Kim and J. O. Rasmussen, Nucl. Phys. {\bf 47}, 184 (1963).
\bibitem{Mello63} P. A. Mello and J. Flores, Nucl. Phys. {\bf 47}, 177 (1963).
\bibitem{Hughes66} T. A. Hughes, R. Snow, and W. T. Pinkston, Nucl. Phys. {\bf 82}, 129 (1966).
\bibitem{Hamada62} T. Hamada and I. D. Johnston, Nucl. Phys. {\bf 34}, 
382 (1962).
\bibitem{Herling72} G.  H. Herling and T. T. S. Kuo, Nucl. Phys. A {\bf 181}, 113 (1972).
\bibitem{Warburton91} E. K. Warburton and B. A. Brown, Phys. Rev. C {\bf 43}, 602 (1991).
\bibitem{Ma73} C. W. Ma and W. W. True, Phys. Rev. C {\bf 8}, 2313 (1973).
\bibitem{Alexa97a} P. Alexa, J. Kvasil, N. V. Minh,  and R. K. Sheline, Phys. Rev. C {\bf 55}, 179 (1997).
\bibitem{Bogner02} S. Bogner, T. T. S. Kuo, L. Coraggio, A. Covello, and N. Itaco, Phys. Rev.
C {\bf 65}, 051301 (2002).
\bibitem{Kuo90} T. T. S. Kuo and E. Osnes, {\it Lecture Notes in Physics}, 
Vol.~364, (Springer-Verlag, Berlin, 1990).
\bibitem{Coraggio06} L. Coraggio, A. Covello, A. Gargano, and N. Itaco, Phys. 
Rev. C {\bf 73}, 031302(R) (2006).
\bibitem{Machleidt01} R. Machleidt, Phys. Rev. C {\bf 63}, 024001 (2001).
\bibitem{Coraggio05} L. Coraggio, A. Covello, A. Gargano, and N. Itaco, Phys. 
Rev. C {\bf 72}, 057302 (2005).
\bibitem{Covello2}  A. Covello, L. Coraggio, A. Gargano, and N. Itaco,
 Prog. Part. Nucl. Phys. {\bf 59}, 401 (2007).
\bibitem{Shurpin83} J. Shurpin, T. T. S. Kuo, and D. Strottman, Nucl. Phys. A  
{\bf 408}, 310 (1983).
\bibitem{Suzuki80} K. Suzuki and S. Y. Lee, Prog. Theor. Phys. {\bf 64},2091 
(1980).
\bibitem{NNDC} Data extracted using the NNDC On-line Data Service from the 
XUNDL database, file  revised as of July 6, 2007.  
\bibitem{Audi03} G. Audi, A. H. Wapstra, and C. Thibault, Nucl. Phys. A  
{\bf 729}, 337 (2003).
\bibitem{Oxbash} B. A. Brown, A. Etchegoyen, and W. D. M. Rae, The computer 
code OXBASH, MSU-NSCL, Report No. 524.
\bibitem{Covello04} A. Covello, L. Coraggio, A. Gargano, and N. Itaco,
Phys. Atom. Nucl. {\bf 67}, 1611 (2004).

\end{thebibliography}
\end{document}